\documentclass[12pt]{iopart}
\usepackage{graphicx}
\begin{document}

\title[A conjectural experiment to observe the effect of locked-down in an epidemic]{A Conjectural Experiment to Observe the Effect of Conditional locked-down in an Epidemic}

\author{M.E. Hoque and S.K. Das}

\address{Department of Physics, Shahjalal University of Science and Technology\\
Sylhet - 3114, Bangladesh\\}
\ead{mjonyh-phy@sust.edu}
\vspace{10pt}
\begin{indented}
\item[]May 2020
\end{indented}

\begin{abstract}
	In a pandemic like Covid-19, there are many countries of lower-earning cannot provide a complete locked-down within the duration of the detected case. The locked-down may result in famine throughout the region of underdeveloped countries after the outbreak. So, a conjectural setup of an epidemic has been studied by applying specific period of locked-down (30 days) in 5 different scenarios. The stochastic approach to the SEIR (Susceptible, Exposed, Infected and Recovered) model has been used to evaluate the dynamics and the effects of locked-down. It is observed that there exist a suitable period to apply locked-down where more susceptible escape from the infection. The effect of the early (as soon as the infected case detected) and late (with respect to the estimated peak of detected cases for no locked-down) implementation of the locked-down has also been studied and found that the late implementation of locked-down will take the least time to end the epidemic. The CFR (Case Fatality Rate) has also been found to be varied from 7.55 to 8.02 for all the considered scenarios.
\end{abstract}

\vspace{2pc}
\section{Introduction}

In a pandemic like Covid-19 originated at Wuhan, China, the whole world gets stuck all at once for months \cite{live}. The food supply chain gets broken, all the local and global business falls, financial crisis has intensified. To prevent the damages due to the virulent pandemic, all the regional area require to be locked-down. But most of the countries of the world cannot provide food to their people if they get locked-down for months. They have to trade the damages due to the epidemic of their people with foods and economics for the people. Besides, a virus outbreak like coronavirus can take years to control and vaccinated. Thus, it is very necessary to know about the implementation strategy of the locked-down situation to a country of low and middle income.

The SEIR (Susceptible, Exposed, Infected and Recover) compartmental model is a very well standard method in epidemiology to predict the dynamics of the infection \cite{Ande79,May79}. It is well known that the SEIR model can prescribe many important aspects of the dynamics of infectious disease. A number of researches are using this model to analyze the Covid-19 outbreak in the various region along with other models \cite{Engb20,Radu20,Lin20,Lope20,Yang20}. Using this SEIR model, Mwalili \textit{et. al.} suggest that the quarantine and the isolation method will win over the spreading of Coronavirus \cite{Mwal20}. All of those estimations and predictions made by the deterministic model are impressively comply with the situation. These researches also help to make policy and strategy over controlling the epidemic for many countries. 

The stochastic model of the SEIR and its variations are also being using for the prediction of the dynamics of epidemics \cite{Zhon14}. The important part of the stochastic model of epidemics is that it does not require the integration as that of the deterministic model. 
It is also important to consider the communication structure of the individuals for better understanding of the transmission of disease, the effect of social distancing and contact tracing which can be implemented in the stochastic model. An implementation of the network model in stochastic SEIR model has been developed by S. G. Ryan \textit{et. al.} where communication distribution of the individuals can be studied \cite{Ryan20}.

In the case of an epidemic (basic reproduction number is larger than unity), the stochastic SEIR model becomes useful to analyze the implementation of various epidemic control policy. In this work, the authors would like the find out the variation in dynamics due to applying the locked-down (social distancing) in a region. The authors would also like to find out the number of escape population from the epidemic disease. The stochastic network model has been described briefly in the method section. The results of a conjectural experiment have been discussed in the results and discussion section. The summary of the work has been provided in the conclusion section.


\section{Theory}

\begin{figure}[h]
	\centering
	\includegraphics[width=0.8\linewidth]{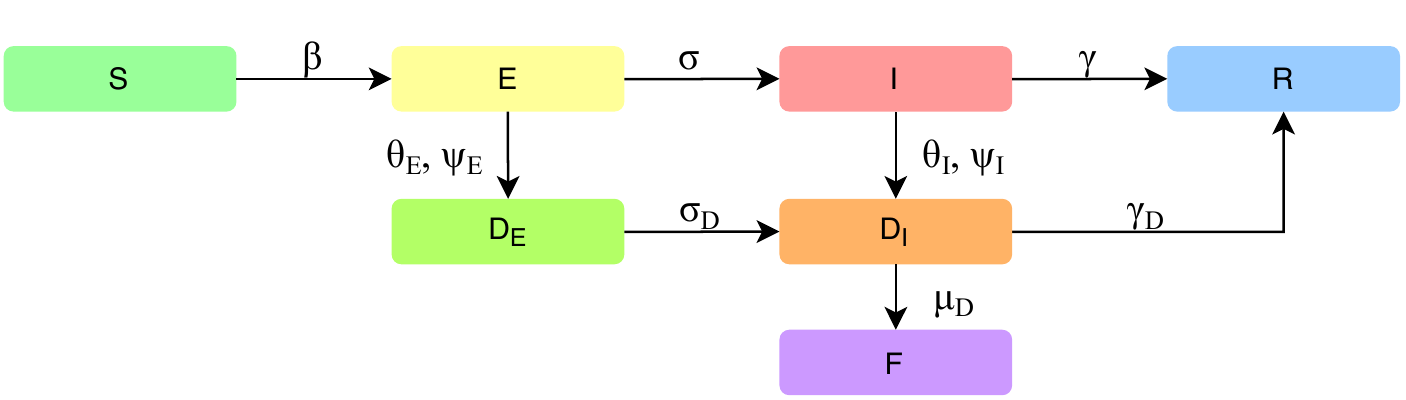}
	\caption{The schematic diagram of the SEIR model where $S,\ E,\ I,\ R,\ D_E$ and $D_I$ represent susceptible, exposed, infected, recovered, detected exposed and detected infected respectively. The arrow represents the transition along with transition rate.}
	\label{fig:hypo_diag_1}
\end{figure}

A schematic diagram of the SEIR compartmental model has been shown in Fig. \ref{fig:hypo_diag_1}. The $S,\ E,\ I,\ D_E,\ D_I,\ R$ and $F$ are the number of susceptible, exposed, infectious, exposed with the detected case, infectious with the detected case, recover and infection-related fatalities for a population of $N$. The $\beta,\ \sigma,\ \gamma,\ \mu_I,\ \beta_D,\ \sigma_D,\ \gamma_D,\ \mu_D,\ \xi,\ \mu_0$ and $\nu$ are the rate of transmission, infection, recovery, infection-related death, transmission for the detected case, infection for the detected case, recovery for the detected case, infection-related death for the detected case, re-susceptibility, baseline death and baseline birth respectively. The $\theta_E,\ \theta_I,\ \phi_E,\ \phi_I,\ \psi_E,\ \psi_I$ and $q$ are the baseline testing for exposed, baseline testing for infectious, testing when a close contact has tested positive for exposed, testing when a close contact has tested positive for an infectious, positive test of exposed state, the positive test of infectious state and the interaction rate of detected infection to the susceptible.

In the network model of population, a graph represents the state of an individual ($S,\ E,\ I,\ D_E,\ D_I,\ R,\ F$) and their interactions. At a given time, each individual interacts with close contacts with a probability of $(1-p)\beta$ and the interaction probability with the global network is $p \beta$.

Each individual $(i)$ in a considered region has a state $X_i$ which updates according to the following probability transition rates \cite{Ryan20} (and Ryan Seamus McGee, personal communication, May 8, 2020):
\begin{equation}
	\label{eq:stoe}
	\fl P(X_i= S \rightarrow E) = \Big[ p \Big(\frac{\beta I + q \beta_D D_I}{N} \Big) + (1-p) \Big(\frac{\beta[\sum_{j\varepsilon C_G(i)} \delta_{X_j=I}] + \beta_D [\sum_{k\varepsilon C_Q(i)} \delta_{X_k = D_I}]}{|C_G(i)|} \Big) \Big] \delta_{X_i = s}
\end{equation}

\begin{equation}
	\label{eq:etoi}
	P(X_i = E \rightarrow I) = \sigma \delta_{X_i = E}
\end{equation}

\begin{equation}
	\label{eq:itor}
	P(X_i = I \rightarrow R) = \gamma \delta_{X_i = I}
\end{equation}

\begin{equation}
	\label{eq:itof}
	P(X_i = I \rightarrow F) = \mu_I \delta_{X_i = I}
\end{equation}

\begin{equation}
	\label{eq:etode}
	P(X_i = E \rightarrow D_E) = \Big( \theta_E + \phi_E \Big[\sum_{j \varepsilon C_G (i)} \delta_{X_k = D_E} + \delta_{X_k = D_I} \Big] \Big) \psi_E \delta_{X_i = E}
\end{equation}

\begin{equation}
	\label{eq:itodi}
	P(X_i = I \rightarrow D_I) = \Big( \theta_I + \phi_I \Big[\sum_{j \varepsilon C_G (i)} \delta_{X_k = D_E} + \delta_{X_k = D_I} \Big] \Big) \psi_I \delta_{X_i = I}
\end{equation}

\begin{equation}
	\label{eq:detodi}
	P(X_i = D_E \rightarrow D_I) = \sigma_D \delta_{X_i = D_E}
\end{equation}

\begin{equation}
	\label{eq:ditor}
	P(X_i = D_I \rightarrow R) = \gamma_D \delta_{X_i = D_I}
\end{equation}

\begin{equation}
	\label{eq:ditof}
	P(X_i = D_I \rightarrow F) = \mu_D \delta_{X_i = D_I}
\end{equation}

\begin{equation}
	\label{eq:atos}
	P(X_i = any \rightarrow S) = \xi \delta_{X_i = R} + \nu \delta_{X_i \neq F}
\end{equation}

With the state probability, $\delta_{X_i = A} = 1$, if the state of $X_i$ is $A$, otherwise it is $0$. The $C_G (i)$ and $C_Q (i)$ denotes the set of close contacts and quarantine contacts of the individual $i$, respectively.

The basic reproduction number, for SEIR model of Covid-19 pandemic  \cite{Engb20}, can be given by,
\begin{equation}
	\label{eq:r0_basic}
	R_0 = \frac{\sigma}{\sigma + \mu_D} \frac{\beta}{\gamma + \mu_D}
\end{equation}

In the case of $\mu_D \rightarrow 0$,
\begin{equation}
	\label{eq:r0}
	R_0 = \frac{\beta}{\gamma}
\end{equation}

which is the expected number of secondary cases produced by a single infected individual in a completely susceptible population. 
\section{Results and Discussions}

The conjectural experiment has been designed for a total number of population $N$, 100000. The magnitude of the parameter $\beta = \beta_D = 1/1.54$, $\sigma = \sigma_D = 1/4.5$ and $\gamma = \gamma_D = 1/3.34$ are chosen in such a way that the detected infection has a width of at least 60 days. 
The other parameters are chosen as $\mu_I = 0.03$, $\mu_D = 0.0004$, $\theta_E = \theta_I = 0.02$, $\phi_E = \phi_I = 0.2$, $\psi_E = \psi_I = 1.0$, $q = 0.5$ and initial infection = 10. In the case of locked-down, the magnitude of $p$ is taken as 0.1 whereas it is 0.5 for the normal situation. These intentional parameters produce 2.17 for the basic reproduction number ($R_0$) (since the $\mu_D$ is very small).

\begin{figure}[h]
	\centering
	\includegraphics[width=0.8\linewidth]{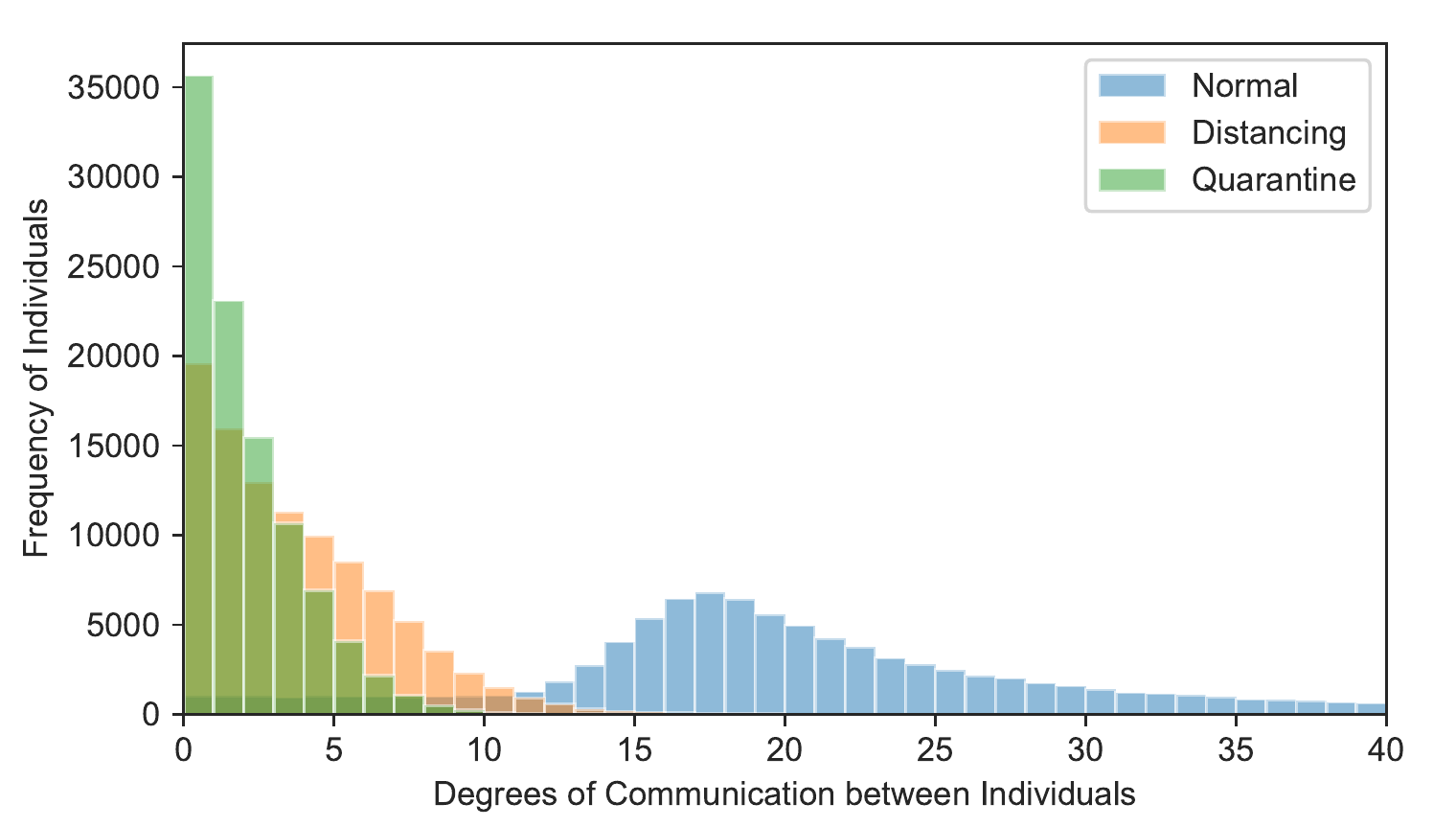}
	\caption{The distribution of the communication of the individuals in a population are shown for normal, locked-down (social distancing) and quarantined situation.}
	\label{fig:hypo_fig_1}
\end{figure}

In this stochastic model, the structured interaction network, Barabasi-Albert Model \cite{Bara99}, has been considered as shown in Fig. \ref{fig:hypo_fig_1}; where the frequency of individuals is plotted against their degrees of communication. The mean degrees of communication for normal, social distancing (locked-down) and quarantine situation is considered as 23.0, 3.3 and 1.6 respectively. The quarantine has been applied to the infected individuals after the detection.

\begin{figure}[h]
	\centering
	\includegraphics[width=0.8\linewidth]{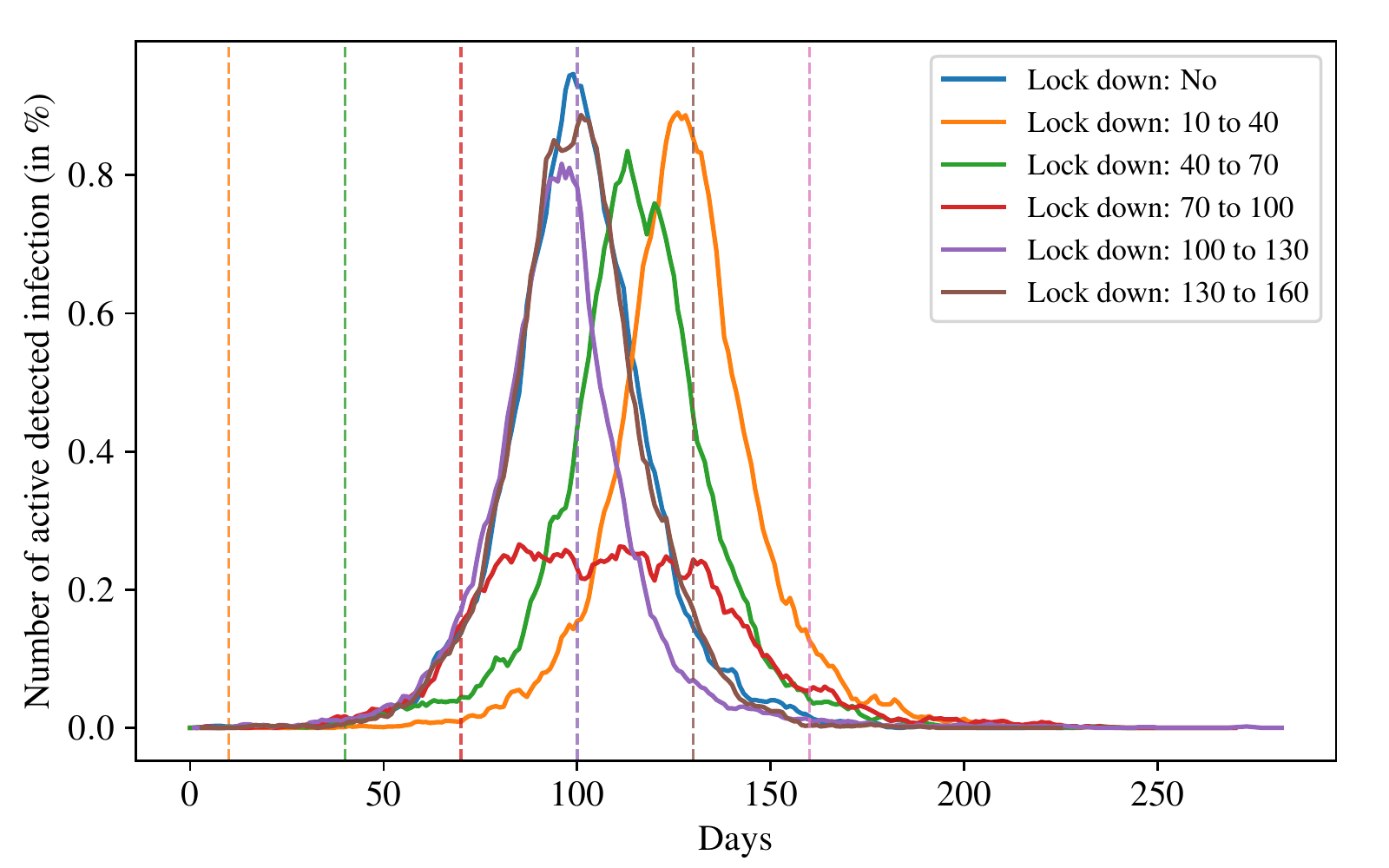}
	\caption{Active detected cases for the 5 of the locked-down scenario including without locked-down. The percentages of the active infected individuals are taken with respect to the total population. The vertical lines indicate the starting of the locked-down for the similar colours of detected cases.}
	\label{fig:hypo_fig_2}
\end{figure}

The values of the parameters are chosen in such a way that the detected case of without locked-down has a width of more than 90 days. It is considered that the epidemic region can be locked-down for only 30 days. Thus, it becomes necessary to apply this 30 days more carefully so that the damage due to epidemic can be minimized. Since the detected case has been found from $30^{th}$ day to $160^{th}$ day in the case of without locked-down scenario (Fig. \ref{fig:hypo_fig_2}), the locked-down phase has been chosen form $10^{th}$ to $40^{th}$,  $40^{th}$ to $70^{th}$,  $70^{th}$ to $100^{th}$,  $100^{th}$ to $130^{th}$ and $130^{th}$ to $160^{th}$ day (say scenario 1 - 5). It is to be noted that the increase in the detected case is found after the $44^{th}$ day. The effect of locked-down for the described scenarios is visualized in Fig. \ref{fig:hypo_fig_2} by considering the detected active cases $(D_E)$. In the case of scenario 1 and 2, the shift in the increase of detected active cases has been observed whereas a steady and decreasing trend has been observed for scenario 3 and 4, respectively. For scenario 5, there is no observable shift whereas the widest shifting is found for scenario 1. It is also observed that the peak of the detected cases for scenario 3 is lower than 0.25\% whereas all other scenarios it crosses the 0.80\% of the total population (percentages are taken with respect to the total population). It indicates that applying conditional locked-down at a suitable time, can control the number of maximum active infected cases. This will help to use the maximum medical facilities which may contribute to minimize the fatal cases.

\begin{figure}[h]
	\centering
	\includegraphics[width=0.8\linewidth]{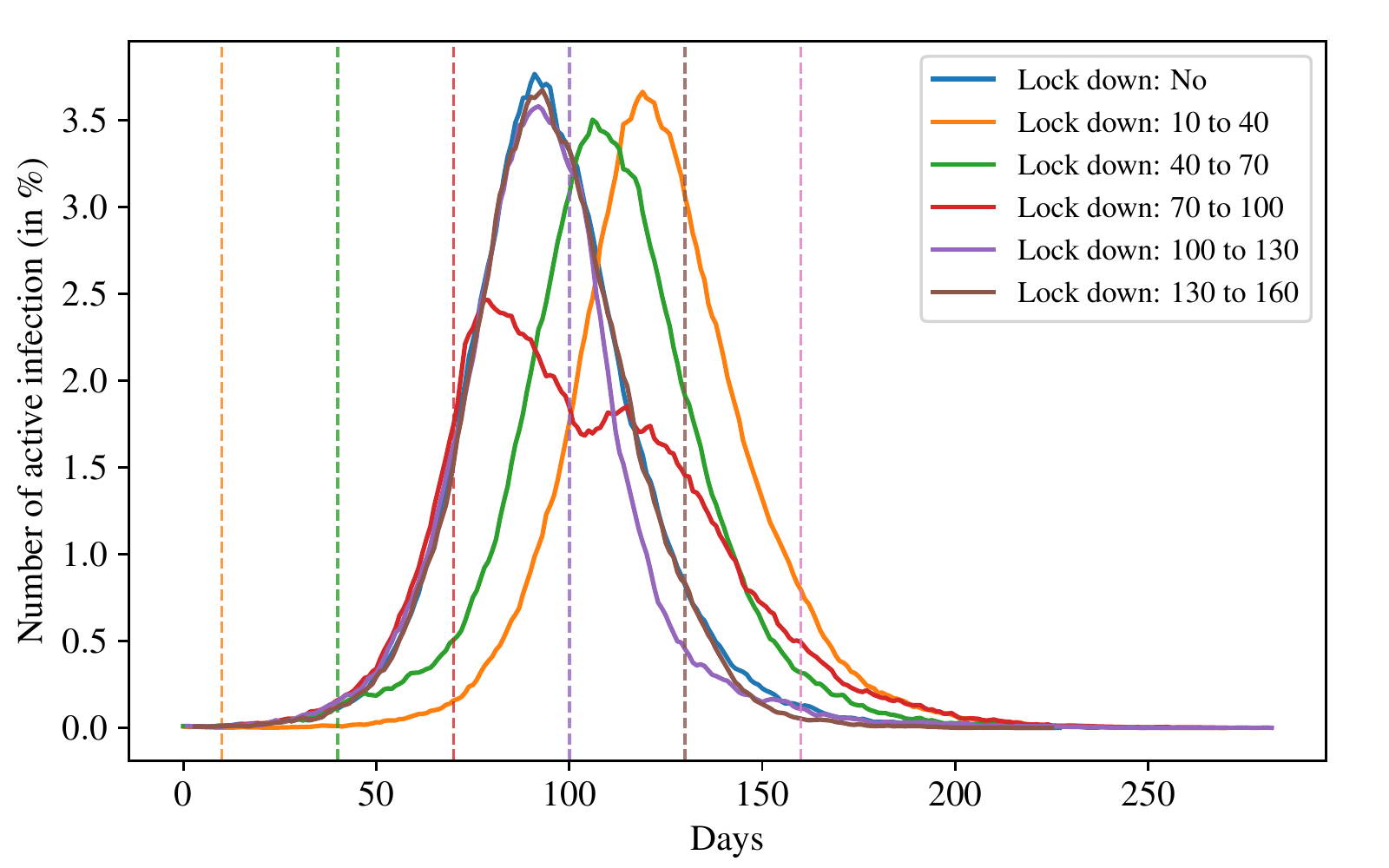}
	\caption{The percentage (with respect to the population) of the active infected case for all the locked-down and no locked-down scenarios are shown.}
	\label{fig:hypo_fig_3}
\end{figure}

The simulated active infected cases ($I$) are shown for all of the scenarios in Fig. \ref{fig:hypo_fig_3}. A similar trend of shifting has been observed as described for Fig. \ref{fig:hypo_fig_2}. But the peak seems to be the same for no locked-down and all the scenarios except 3. In the case of scenario 3, the peak is found after $79^{th}$ day which is 9-days later of the locked-down implementation day $(1/\beta + 1/\sigma + 1/\gamma \approx 9)$. Such a 9-days activity is also found for scenario 4 decrease rapidly after $109^{th}$ day with respect to the no locked-down situation. Since the detected cases are not inter-related (a random fraction of the infected cases are detected), the downward shift of the peak between Fig. \ref{fig:hypo_fig_2} and Fig. \ref{fig:hypo_fig_3} is understandable. All the locked down scenarios have happened as expected for the estimation of infected cases.

\begin{figure}[h]
	\centering
	\includegraphics[width=0.8\linewidth]{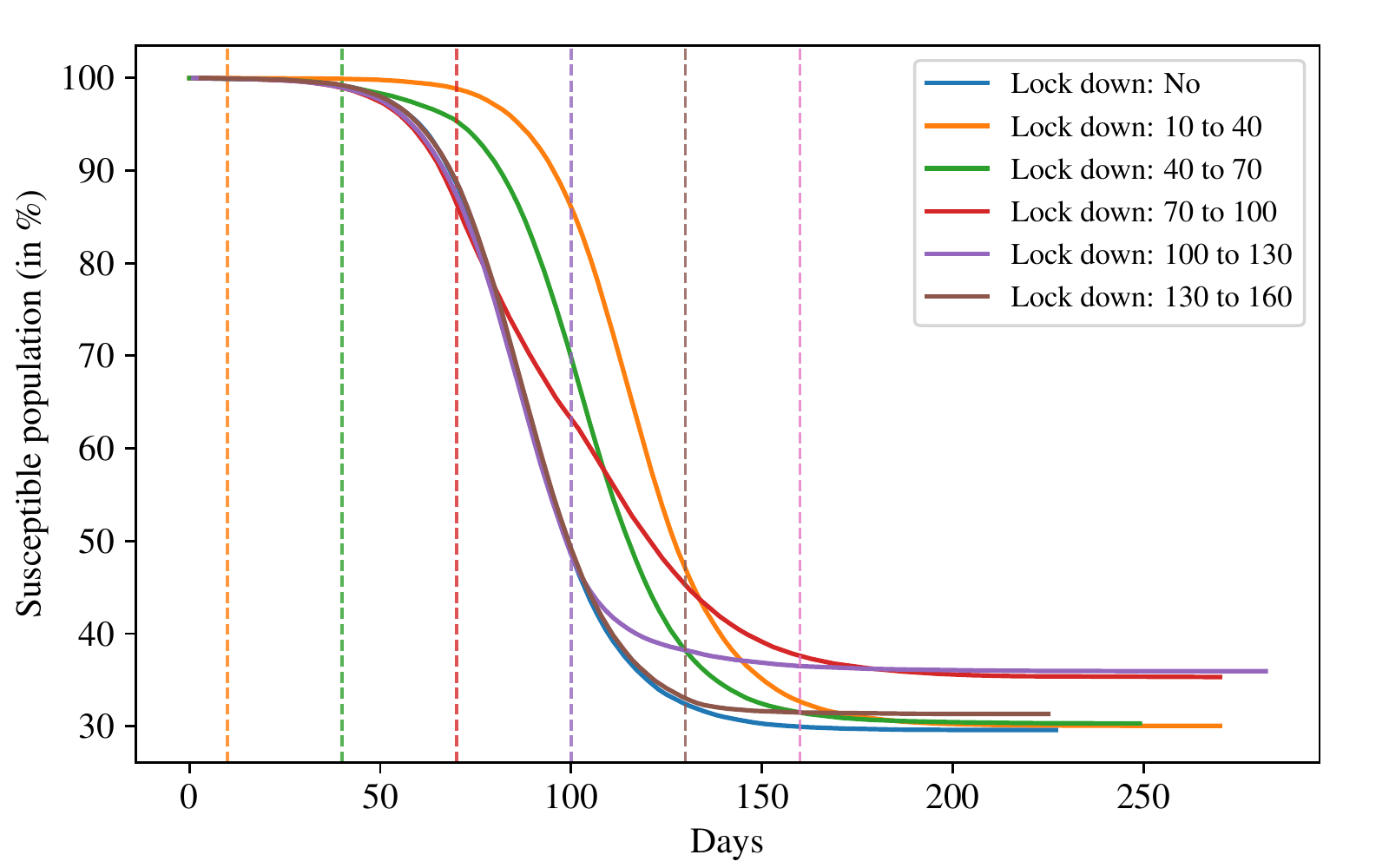}
	\caption{The number of susceptible population (with respect to the total population) are shown in percentage for all the described scenarios.}
	\label{fig:hypo_fig_4}
\end{figure}

The dynamics of the susceptible population ($S$) is the greater interest of this conjectural experiment. Fig. \ref{fig:hypo_fig_4} shows the variation in the dynamics of the susceptible population for all the cases. It is found that there are more than 30\% susceptible population remains unaffected after the epidemic for no locked-down, early and lately locked-down (scenario 1, 2 and 3). It is also observed that this percentage become more than 40\% for scenario 3 and 4 (where locked-down happened near the peak of without locked down). The early or late locked down seems to produce a similar number of unaffected susceptible. These analyses indicate that there exists a suitable time for applying conditional locked down where more people can escape from the infection.

\begin{figure}[h]
	\centering
	\includegraphics[width=0.8\linewidth]{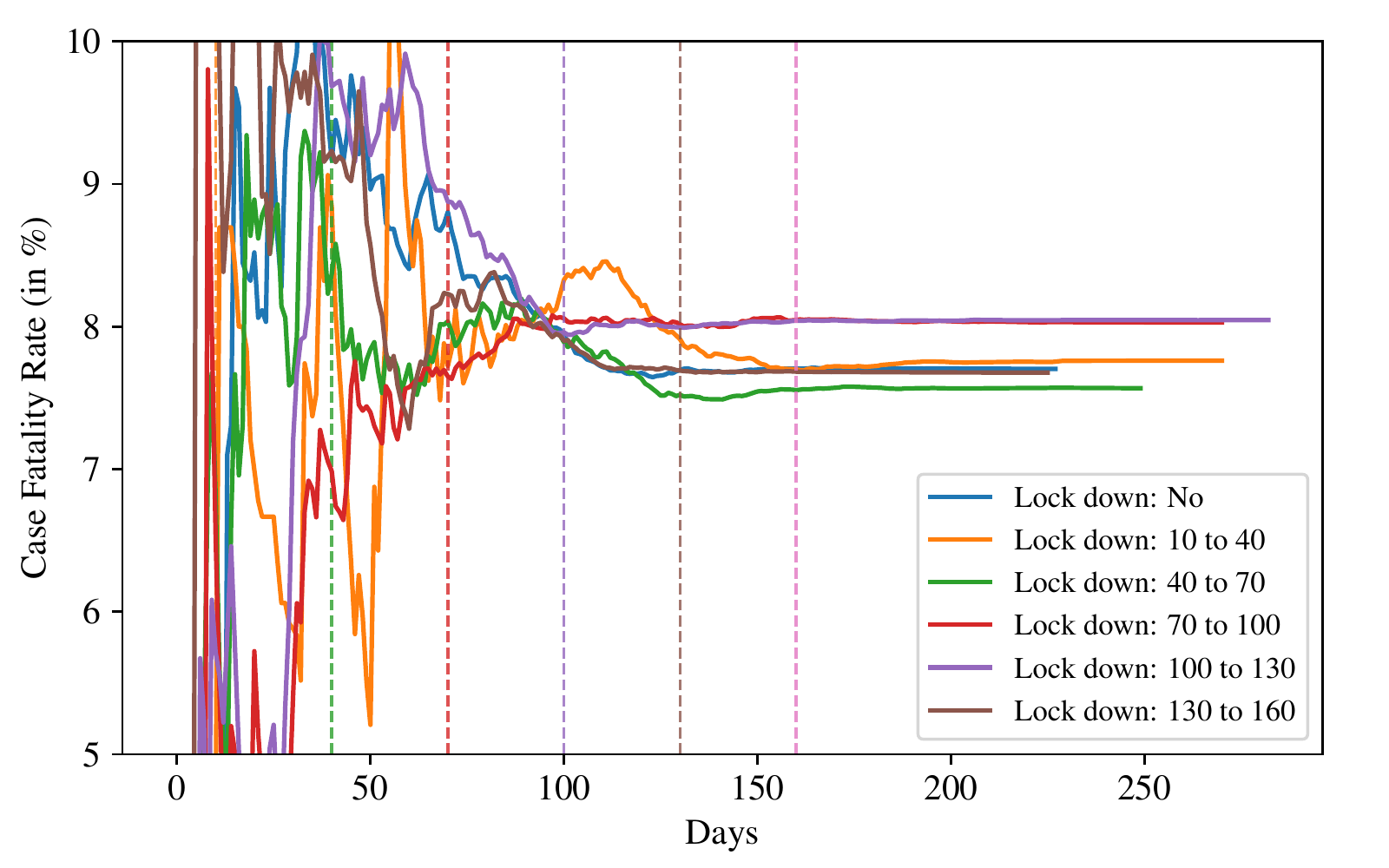}
	\caption{Estimation of the Case Fatality Rate (CFR) for various locked-down scenario.}
	\label{fig:hypo_fig_5}
\end{figure}

The Case Fatality Rate (CFR) calculation is complicated as it should not be \cite{Baud20,Rajg20,Spyc20}. In this article, the CFR has been calculated as the rate of deceased for the recovered population at the same time \cite{Wu20}. It is observed that the CFR varies from 7.55 to 8.02 for all the locked-down scenario including no locked-down case after 130 days (Fig. \ref{fig:hypo_fig_5}). The lowest CFR is found for the scenario 2 whereas it is expected for the scenario 3 because of the steady observation in detected cases (with respect to Fig. \ref{fig:hypo_fig_2}). The initial fluctuations are well understood for their detection and treatment to the patient. The observed under or over estimation in the initial CFR has been well discussed by Rajgor \textit{et.al.} \cite{Rajg20}. Though the number of case fatality depends on many of the other factors (like age, immunity, previous medical condition etc.), the variation in the observed CFR indicates that the fatality can also be depended on the applying conditional locked-down. 

This simulation has been run until no infected cases for the 100000 population. It is observed that the longest and shortest epidemic occurs for scenarios 1 and 5. Which describes whether to chose longest or shortest time.

\section{Conclusions}

A conjectural experiment has been designed to predict the effect of locked-down in an epidemic. The Barabasi-Albert communication network model has been used to produce deploy the regional normal, social distancing and quarantine communication of individuals. The stochastic SEIR model has been intentionally parameterized so that the detected active case can be observed for more than 90 days. A 30 days locked-down period has been applied in 5 of the specific scenarios. The above mentioned notional experiment with intentional parameters can be summaries as:
\begin{enumerate}
	\item If the allocated 30 days locked-down period has been applied too early (scenario 1 and 2) or lately (scenario 5), it barely modifies the dynamics of no locked-down case.
	\item In the case of scenario 1 and 2, the dynamics only shifted for more than 30 and 20 days respectively. So, the locked-down should not be applied too early, if it requires to maintain a conditional locked-down.
	\item In the study of the dynamics of susceptible population, interestingly scenario 3 and 4 has more than 40\% susceptible individual remains after the outbreak.
	\item It is observed that the initial fluctuation of CFR becomes steady after the $93^{rd}$ day (the day where the peak occurs in the infected case due to no locked-down condition Fig. \ref{fig:hypo_fig_3}). The lowest CFR is 7.55 which is found for scenario 2.
	\item Very interestingly, scenario 5 (where late locked-down has been introduced) takes least time to reach the end of the epidemic. And the scenario 1 takes the longest time to reach at the end.
\end{enumerate}

Though the verification of the above mentioned experiment is not feasible in reality, the authors believe that the speculated summary will impact very efficiently to make the policy for controlling the epidemic.

\section*{Acknowledgements}%
\label{sec:acknowledgements}

The authors would like to thank Prof. Dr. Zafar Iqbal, Department of Computer Science and Engineering, Shahjalal University of Science and Technology, Sylhet - 3114, Bangladesh for his valuable discussions.
\section*{References}%
\label{sec:references}

\end{document}